\documentclass{ws-ijmpcs}

\newcommand{\mnras}{{\it MNRAS }}
\newcommand{\apj}{\it {ApJ }}
\newcommand{\apjl}{{\it ApJ }}

\newcommand{\apss}{{\it Ap\&SS }}

\newcommand{\swift}{{\it Swift}}

\begin{document}

\markboth{P.A. Curran}
{Observations of particle acceleration in the blast waves of Gamma-ray Bursts}

%
\catchline{}{}{}{}{}
%

\title{Observations of particle acceleration in the blast waves of \\Gamma-ray Bursts}

\author{Peter A.~Curran}

\address{Laboratoire AIM, CEA DSM/IRFU/SAp, Centre de Saclay, F-91191  Gif-sur-Yvette, France\\
peter.curran@cea.fr}

\maketitle

\begin{history}
\received{Day Month Year}
\revised{Day Month Year}
\end{history}

\begin{abstract}
The electron energy distribution index, $p$, is a fundamental parameter of the process by which electrons at shock fronts are accelerated to relativistic speeds and by which they radiate, via synchrotron emission. This acceleration process is applicable to a myriad of astronomical jet sources such as AGN, X-ray binaries and gamma-ray bursts (GRBs), as well as to particle acceleration in the solar wind and supernovae, and the acceleration of cosmic rays. The accurate measurement of the distribution of $p$ is of fundamental importance to differentiate between the possible theories of electron acceleration at any relativistic shock front; there is division as to whether $p$ has a universal value or whether it has a distribution, and if so, what that distribution is.

Here one such source of synchrotron emission is examined: the blast waves of GRB afterglows observed by the \swift\ satellite. Within the framework of the GRB blast wave model, the constraints placed on the distribution of $p$ by the observed X-ray spectral and temporal indices are examined and the distribution parametrized.  The results show that the observed distribution of spectral indices is inconsistent with an underlying distribution of $p$ composed of a single discrete value but consistent with a Gaussian distribution centred at $p = 2.4$ and having a width of $0.6$. This finding disagrees with theoretical work that argues for a single, universal value of $p$, but also demonstrates that the width of the distribution is not as wide as has been suggested by some authors. 

\end{abstract}

\ccode{PACS numbers: 95.75.Pq, 95.85.Nv, 98.70.Rz}

\section{Introduction}

The afterglow emission of Gamma-Ray Bursts (GRBs) is generally accepted to originate from external shocks when a collimated ultra-relativistic jet ploughs into the circumburst medium, driving a shock front ahead of it\cite{rees1992:mnras258,meszaros1998:apj499}. The shock accelerates charged particles in the jet or ambient medium to the relativistic speeds required so that they may radiate electromagnetically via synchrotron emission. 
This acceleration mechanism -- common to many astronomical jet sources such as AGN, X-ray binaries and GRBs (as well as particle acceleration in the solar wind and supernovae, and the acceleration of cosmic rays) -- is thought to be Fermi diffusive shock acceleration\cite{fermi1954:ApJ119} due to the passage of an external shock\cite{blandford1978:ApJ221,rieger2007:Ap&SS309} after which the energy of the electrons, $E$, follows a power-law distribution, $N(E) \mathrm{d}E \propto E^{-p} \mathrm{d}E$, with a cut-off at low energies. Fundamental to this is the electron energy distribution index, $p$, the characteristic parameter of the process,
which is dependent only on the underlying micro-physics of the acceleration process. Some (semi-)analytical calculations and simulations indicate that there is a nearly universal value of $\sim 2.2-2.4$\cite{achterberg2001:MNRAS328,spitkovsky2008:ApJ682,kirk2000:ApJ542} though other studies suggest that there is a large range of possible values for $p$ of $1.5-4$\cite{baring2004:NuPhS136}. 
Observationally, different methods have been applied to samples of BATSE, {\it BeppoSAX} and \swift\ GRBs which reached the conclusion that the observed range of $p$ values is not consistent with a single central value of $p$\cite{chevalier2000:ApJ536,panaitescu2002:ApJ571,shen2006:MNRAS371,starling2008:ApJ672,curran2009:MNRAS395} and the latter three showed that the width of the parent distribution is $\sigma_{p} \sim 0.3-0.5$. 
In these proceedings I  summarize our previous work\cite{Curran2010:ApJ716L.135C} in which we interpret a much larger and, statistically, more significant sample of \swift\ observed GRB afterglows, to constrain the electron energy distribution, $p$. I further compare the distributions found from spectral and temporal indices, as well as those found for other sources.

\section{Method and Results}\label{section:method}

\begin{figure}
\centerline{\psfig{file=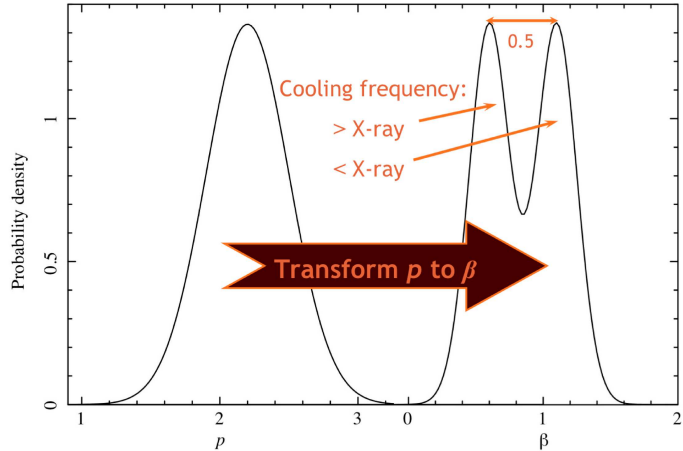,angle=-90,width=9cm}}
\caption{The left curve shows the probability distribution function of the underlying value of $p$, while the right, double-peaked curve shows the transformation of that into  X-ray spectral index space. The distance between the peaks is equal to the difference between the spectral indices either side of a cooling break ($\Delta\beta = 0.5$), while the relative heights of the two peaks are dependent on the relative probability that the X-ray regime falls above or below the cooling frequency. \label{f1}}
\end{figure}

The  method,  detailed in Ref. \refcite{Curran2010:ApJ716L.135C}, is to constrain the electron energy distribution index, $p$, from the values of the X-ray spectral indices, $\beta_{{\rm X}}$ or $\beta$,
observed by the \swift\ XRT up to July 2008\cite{evans2009:MNRAS397}.
$p$ is derived from the spectral index as opposed to the temporal index because for a given spectral index, assuming the asymptotic limit, there are only two possible values of $p$ depending on whether the cooling frequency, $\nu_{{\rm c}}$, is less than or greater than the X-ray regime, $\nu_{{\rm X}}$, while for a given temporal index there are multiple possible values which are model dependent (e.g., the simple  blast wave model\cite{rees1992:mnras258,meszaros1998:apj499}, or modifications thereof).
In accordance with synchrotron emission predicted by the blast wave model, the behaviour of the unabsorbed X-ray spectrum is described as a single power law, where the flux goes as $F_{\nu}(\nu) \propto  \nu^{-\beta}$.
Under the standard assumptions of slow cooling and adiabatic expansion of the blast wave, the electron energy distribution index is related, in the asymptotic limit, to the spectral index by either $p = 2\beta$ ($\nu_{\mathrm{c}} < \nu_{\mathrm{X}}$) or $p = 2\beta +1$ ($\nu_{\mathrm{c}} > \nu_{\mathrm{X}}$) (e.g., Ref. \refcite{granot2002:ApJ568});  implying a difference between the slopes of the two spectral regimes of $\Delta\beta = 0.5$.  
The regime probability, $X$, is defined as the probability that the cooling frequency is less than the frequency of the X-ray regime (i.e., $\nu_{\mathrm{c}} < \nu_{\mathrm{X}}$) and $1-X$ is the probability that $\nu_{\mathrm{c}} > \nu_{\mathrm{X}}$  (Figure \ref{f1}). The case where the cooling frequency may be passing through the X-ray regime is neglected (since there is no sign of spectral evolution in the sample) as are  the cases where the peak frequency, $\nu_{\mathrm{m}}$, or self absorption frequency, $\nu_{\mathrm{a}}$, are greater than the X-ray regime as these are not observed in late time afterglows.  
To parameterise the underlying distribution of the electron energy distribution index, $p$, a maximum likelihood fit is used to return the \emph{most likely} parameters of the assumed underlying model, the errors on which are estimated via a Monte Carlo error analysis. Another Monte Carlo analysis tests the probability that the observed distribution of spectral indices could be obtained from an underlying distribution of $p$ described by the most likely parameters.

\begin{figure}
\centerline{\psfig{file=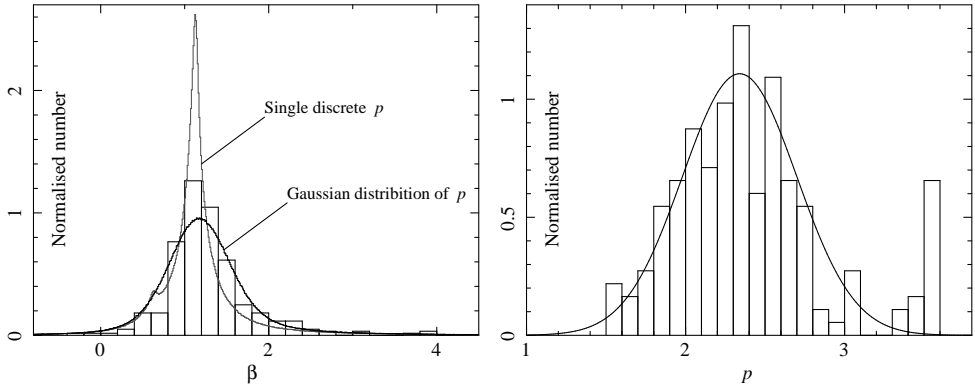,angle=-90,width=13cm}}
\caption{{\it Left:}
Normalised histogram of the spectral data (301 measurements) overlaid with high-resolution normalised histograms of the synthesized data sets ($10^4 \times 301$ data points) from the most likely parameters of a single discrete $p$ (narrow distribution) and a Gaussian distribution of $p$ (wide distribution). Note that there is only one obvious peak on the spectral index distribution while from the transformation of figure \ref{f1}, two were expected. 
{\it Right:} Normalised histogram of the 191 values of $p$ published in Ref. \protect\refcite{evans2010:sif102}, overlaid with the probability density function of $p = 2.36 $ and $\sigma = 0.36$, which is consistent with that data.  
\label{f2}}
\end{figure}

There are two underlying models, or hypotheses, regarding the data that we wanted to test: that the observed distribution of spectral indices, $\beta$, can be obtained from an underlying distribution of $p$ composed of $i)$ a single discrete value and $ii)$ a Gaussian distribution.  
We found\cite{Curran2010:ApJ716L.135C} that the observed distribution of spectral indices are inconsistent with the first hypothesis but consistent with the second, a Gaussian distribution centred at $p = 2.36$ and having a width of $0.59$ (Figure \ref{f2} {\it Left}). Furthermore, if one accepts that the underlying distribution is a Gaussian, the majority ($\gtrsim 94\%$) of GRB afterglows in our sample have a cooling break frequency less than the X-ray frequency.

\section{Discussion}

To further test the conclusion regarding the distribution of $p$, the values of $p$ published for 191 GRBs\cite{evans2010:sif102} are tested in a similar way to above using  Monte Carlo analyses. It is found that these data (Figure \ref{f2} {\it Right}) are well fit by a Gaussian distribution of $p$, centered at $p = 2.36 $ and with a width of $\sigma = 0.36$. It is important to note that this is a totally independent data set from that described above -- based on temporal fits of XRT light curves as opposed to spectral fits -- and as such offers an independent test of our results for a similar \swift-XRT GRB sample.
The excellent agreement between these two independent data sets offers confirmation of the results for the distribution of $p$, as well as our conclusion that the majority of GRB afterglows in the sample have a cooling break frequency less than the X-ray frequency, since both are interdependent in our study.

This result confirms the results from previous small-sample GRB afterglow studies\cite{shen2006:MNRAS371,starling2008:ApJ672,curran2009:MNRAS395} as regards the non-universality of $p$, the central value at $p \sim 2.0 - 2.5$, and the width of the distribution of $\sigma_{p}\sim 0.3 - 0.5$.  However, our results are based on a sample of bursts an order of magnitude larger than these studies and a statistical approach to the position of cooling frequency relative to the X-ray is taken, by using the regime probability, $X$. 
The result is also consistent with (semi-)analytical calculations and simulations that indicate a value of $p \sim 2.2-2.4$\cite{kirk2000:ApJ542,achterberg2001:MNRAS328,spitkovsky2008:ApJ682}. Though it rules out the suggestion of a single, universal value, the result demonstrates that the distribution is not as wide as has been suggested by some authors ($1.5 \lesssim p \lesssim 4$)\cite{baring2004:NuPhS136}.

Given that these results suggest  the value of $p$ is not universal, it may be possible that $p$ changes suddenly or evolves gradually with time or radius even in a single event as environmental shock parameters (e.g., magnetic field, ambient density) change or evolve. It is also possible that different components of a structured jet, multi-component jet or jet-cocoon could have different values of $p$. If either are the case our derived values of $p$ should be considered time averaged values of the parameter, though a change or evolution of $p$ 
should be observable as a change or evolution of the synchrotron spectral index, $\beta$, and no significant example of such an evolution has been observed in GRB afterglows. \\


It is very clear from the data (Figure \ref{f2} {\it Left}), as well as the fitting, that there is no secondary peak as had been expected (Figure \ref{f1}) and this implies that the majority of GRB afterglows in the sample ($\gtrsim 94\%$) have a cooling break frequency less than the X-ray frequency. While we found a distribution of $p$ that concurs with previous work, this is somewhat anomalous, as previous multi-band (optical -- X-ray) studies\cite{panaitescu2002:ApJ571,starling2008:ApJ672,curran2009:MNRAS395} have shown that the cooling break frequencies of a number of GRB  afterglows are greater than the X-ray frequency. 
The upper limit  derived implies that this is true in only $\sim20$ of 301 GRBs in the sample and  while this is a discrepancy from the ratio observed in the multi-band studies (5 out of 10\cite{starling2008:ApJ672} and 2 out of 6\cite{curran2009:MNRAS395}) it is not seriously so, given the extremely low number statistics of those studies. In fact, if  a sample of 6 random bursts is created from the statistical distribution of 301, 
there is a non-negligible ($\sim$ 5\%) chance that 2, or more, 
bursts would have a cooling frequency greater than the  X-ray frequency, 
consistent with the aforementioned \emph{Swift} based study\cite{curran2009:MNRAS395}, which investigates the same population.

The lack of a second peak in the data, if not intrinsic, could be due to a very wide underlying distribution of $p$ which would blur out the double peak structure; however this requires a far wider distribution than is observed. It might also be explained by a bimodal distribution of $p$ where the mode from which $p$ is  drawn is correlated with position of cooling break but this is a highly convoluted and unlikely scenario. Another possible explanation is that instead of $p$ being related to the spectral index through one of the two relationships presented in section \ref{section:method}, there is only a single relationship or a continuous range of relationships. This is certainly possible, most obviously caused by the fact that the canonical GRB spectra does not have sharp breaks between the regimes but is, instead, smoothly transitioning\cite{granot2002:ApJ568}. This would cause a continuous range of relationships which would manifest as a single peaked spectral index distribution centered about the average of the two relationships, $\beta \approx (p-0.5)/2$, and which would, in the case of our data imply that $p \approx 2.9$ with a width of $\sigma_{p} \approx 0.6$.  However, importantly the cooling frequency evolves in time so any such spectra due to a smooth transition between regimes would likewise evolve and no such significant evolution has been observed in these GRB afterglows\cite{evans2009:MNRAS397}, leading us to conclude that this is unlikely to be an effect and that the single peak structure is, in fact, intrinsic.

\subsection{Comparison with other sources}

A similar, though smaller sample, effort has been made to derive the distribution of $p$ in blazars and pulsar wind nebulae\cite{shen2006:MNRAS371}  which found that  $p \approx 2.4$ ($\sigma_p \approx 0.4$) and  $p \approx 1.7$ ($\sigma_p \approx 0.6$) respectively. But the uncertainty of these authors' methods is demonstrated by the fact that it was also possible that the distribution could be described as $p \approx 3.2$ ($\sigma_p \approx 0.2$) and $p \approx 2.0$ ($\sigma_p \approx 0.2$) respectively, when the distribution was forced to be as narrow as possible.
Though, no such attempt has been made to statistically analyze X-ray binaries, the observed radio spectra of these sources have spectral indices of order 0.5-0.7, corresponding to  $p \approx$ 2.0-2.3, while a small sample of FR I active galaxies have been treated to a non-statistical analysis\cite{Young2005:ApJ.626} which found a narrow range at $2.1$. 
It is planned to expand the above described statistical method to these other sources to obtain accurate measurements of the distribution in the different classes. These may then be used  as a  touchstone for theories and simulations and will allow  proposed models to be tested.

\section{Conclusions}

These results for \swift\ observed GRBs show that the observed distribution of spectral indices are inconsistent with an underlying distribution of $p$ composed of a single discrete value but consistent with a Gaussian distribution centred at $p = 2.4$ and having a width of $0.6$. This finding disagrees with the theoretical work that argues for a single, universal value of $p$, but also demonstrates that the width of the distribution is not as wide as has been suggested by some authors. Due to the lack of accurately measured distributions of $p$ for other sources, an accurate comparison with those sources is not possible, however, it is planned to expand this statistical method to  other sources to obtain accurate measurements of the distribution in the different classes, which may then be used  as a  touchstone for theories and simulations.

\section*{Acknowledgments}

The author thanks Phil A. Evans for useful discussions. 
PAC is supported by the Centre National d'Etudes Spatiales (CNES), France through MINE: the Multi-wavelength INTEGRAL NEtwork.


\end{document}